\documentclass[letterpaper]{article}
\usepackage{spconf}
\usepackage{cite}
\usepackage{amsmath,amssymb,amsfonts}
\usepackage{algorithmic}
\usepackage{graphicx}
\usepackage{textcomp}
\usepackage{xcolor}
\usepackage{booktabs}
\usepackage{multirow}
\usepackage{pifont}
\usepackage{siunitx}
\usepackage{enumitem}
\usepackage[protrusion=false]{microtype}
\usepackage{flushend}
\usepackage[linesnumbered,ruled,vlined]{algorithm2e}
\usepackage[hidelinks]{hyperref}
\SetAlgoNlRelativeSize{0}

\title{VISH-ASR: Viterbi-guided Selection of Hotwords for Speech-LLM Contextual ASR}
\name{Xiaoyu Gu$^{1}$, Yu Xi$^{1}$, Chencheng Wang$^{1}$, Zonghan Zhou$^{1}$, Haoyu Li$^{2}$, Jing Peng$^{1}$, Kai Yu$^{1}$\sthanks{Kai Yu is the corresponding author.}}
\address{$^{1}$MoE Key Lab of Artificial Intelligence, AI Institute, X-LANCE Lab, Shanghai Jiao Tong University\\
$^{2}$School of Intelligence Science and Technology, Nanjing University, Suzhou, China}

\begin{document}
\ninept

\maketitle

\begin{abstract}
Prompt-based contextual ASR requires selecting a compact hotword subset
from large dynamic inventories, yet existing pre-selection methods rely either on discrete ASR hypotheses that may discard acoustic evidence or on
separately trained retrieval models. We propose
Viterbi-guided Selection of Hotwords (VISH), a training-free,
plug-and-play framework that directly verifies candidate-specific
acoustic support over frame-level CTC posteriors. VISH formulates
pre-selection as constrained posterior search; M-VISH makes large-list
verification scalable through frame-synchronous shared-prefix Viterbi
search with bounded top-$K$ pruning, while Bi-MVISH exploits
complementary backward evidence for robust verification. Across five benchmarks, M-VISH achieves up to 11.7\% absolute improvement in Best-F1 over a strong learned-retrieval baseline and reduces candidate-search latency by 75.2\% at 2,000 candidates relative to hypothesis-based filtering. These results demonstrate the effectiveness of posterior-level verification for accurate and scalable hotword pre-selection without additional model training.
\end{abstract}


\section{Introduction}
\label{sec:intro}

End-to-end automatic speech recognition (ASR) performs well on
general-domain corpora~\cite{li2022recent,ji2024wavchat}, yet remains
vulnerable to hotwords: semantically critical long-tail terms, such as
contact names, locations, and products, that are sparsely represented in
training data. These user- and task-specific terms are better supplied as
external lexical context~\cite{shi2024seaco,hori2018end,chen2019end}.
Contextual ASR has used weighted finite-state transducers
(WFSTs)~\cite{DBLP:conf/interspeech/BakhturinaZG22} and neural context
injection~\cite{pundak2018deep,DBLP:conf/interspeech/JainKMZMS20,
variani2020hybrid,zhao2019shallow}. Speech large language models
(Speech LLMs) offer a new paradigm~\cite{song2023contextual,
asano2025contextual,peng2025survey}.

Speech-LLM contextualization follows two paradigms. Architecture
modification or extensive fine-tuning learns hotword-aware
representations~\cite{fu2025pac,ren2026rlbr,jing2025rare,
kong2025contextual}, but is computationally costly, inflexible for
dynamic contexts, and difficult to generalize without retraining.
Alternatively, prompt-based systems inject hotwords directly into the
input without adapting the Speech LLM~\cite{yang2023two,chen2024salm,
gong2025br}, preserving model generality and enabling inexpensive
per-request personalization. However, large candidate lists weaken prompt
conditioning, increase interference, and cause bias-driven substitutions in
both hotwords and background words~\cite{gong2025br}. Most systems
therefore pre-select a compact candidate subset that retains acoustically
supported hotwords while rejecting irrelevant
alternatives~\cite{yang2023two}.

Existing pre-selection methods differ mainly in how candidate evidence is
obtained. Hypothesis-based filtering relies on discrete recognition
outputs and candidate matching~\cite{chen2019joint,yang2024ctc,
asano2025contextual}. For example, CTC-Assistant~\cite{yang2024ctc}
first decodes a CTC hypothesis and then filters hotwords through lexical
indexing and edit-distance matching; relevant acoustic evidence may
therefore be lost when the candidate is absent from the discrete
hypothesis. Learned speech--text retrievers, including BR-ASR, GLCLAP,
and CLAR, require separately trained retrieval
models~\cite{gong2025br,kong25_interspeech,huang2026clar}.
Search-graph approaches based on WFST composition or lattice
rescoring~\cite{hori2007efficient,wang2023contextual} often require
decoder-side graphs, lattices, or task-specific search structures, making
them less convenient for black-box contextual ASR. CTC-based keyword
spotting has also explored posterior-level keyword-constrained alignment
for detection~\cite{xi2025streaming}; however, its objective is keyword
detection rather than ranking and selecting a compact subset from a large,
dynamically changing inventory for downstream contextual prompting.

We propose Viterbi-guided Selection of Hotwords (VISH), which formulates
large-list hotword pre-selection as candidate-specific constrained search
directly over frame-level CTC posterior trajectories. Unlike
hypothesis-based filtering, VISH does not require a candidate to appear in
a discrete first-pass transcript, and it requires no additional
retrieval-model training. To make this formulation scalable,
Multi-Word VISH (M-VISH) introduces a frame-synchronous shared-prefix
search that reuses constrained-search computation among candidates with
common token prefixes and bounds the active search space through
top-$K$ pruning. Bidirectional M-VISH (Bi-MVISH) further exploits
complementary backward acoustic evidence to improve verification robustness.

The contributions of this work are threefold:
\begin{enumerate}
    \item We formulate large-list hotword pre-selection as
    candidate-specific constrained search over frame-level CTC
    posteriors and propose VISH, enabling direct acoustic
    verification without relying on discrete ASR hypotheses
    or additional retrieval-model training.

    \item We introduce M-VISH, a frame-synchronous shared-prefix
    multi-candidate search with bounded top-$K$ pruning, making
    acoustic verification scalable to large dynamic inventories.
    Bi-MVISH further exploits complementary backward evidence
    for more robust verification.

    \item Experiments across multiple datasets and ASR systems
    demonstrate that improved pre-selection translates into
    lower downstream biased error while preserving overall
    recognition accuracy, with both native and external CTC
    posteriors supporting flexible integration.
\end{enumerate}

\begin{figure*}[t]
    \centering
    \includegraphics[width=1\linewidth]{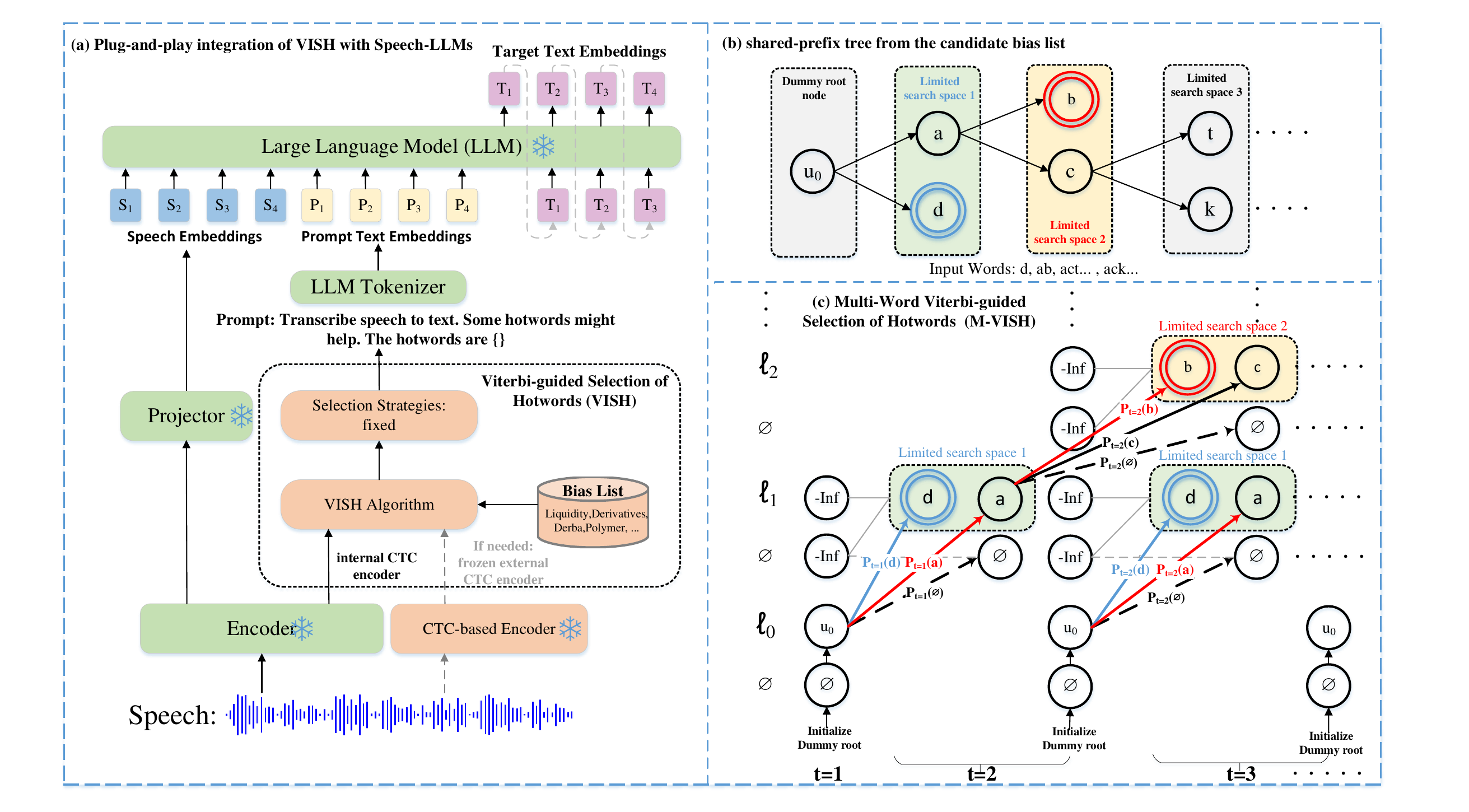}
    
    \vspace{-10pt}
\caption{
Overview of VISH and the M-VISH search procedure.
(a) VISH verifies candidate hotwords using internal or external CTC
posteriors and selects a compact subset for contextual prompting
without modifying the downstream ASR model.
(b) A shared-prefix trie organizes candidate CTC-token sequences
to reuse Viterbi computations across candidates.
(c) At each retained frame, M-VISH updates constrained Viterbi
scores, retains the top-$K_{\mathrm{p}}$ active prefixes, and records
scores at terminal nodes for scalable large-list verification.
Here, $p_t(k)$ denotes the CTC posterior of token $k$ at frame $t$,
and $\ell$ denotes the trie depth.
}
    \label{fig:stucture}
    \vspace{-10pt}
\end{figure*}

\section{Methodology}

\subsection{A plug-and-play VISH framework}
\label{sec:plug_and_play}

Given speech and a candidate list, VISH estimates candidate-specific acoustic support from frame-level CTC posteriors obtained from an accessible internal encoder or a frozen external CTC scorer (Figure~\ref{fig:stucture}). Here, ``training-free'' means no additional task-specific or retrieval-model training or fine-tuning: an off-the-shelf pretrained CTC scorer is used frozen for verification. An external scorer adds acoustic-encoding cost when internal posteriors are unavailable. VISH returns a compact prompt subset without downstream-model modification or decoder access. Verification relies on reliable candidate-token posterior evidence; unusual pronunciations or spelling variants may weaken the score.

\subsection{Multi-word Viterbi-guided selection of hotwords}
\label{sec:m-v-h-s}

Let $p_t(k)=P(\pi_t=k\mid\mathbf{x})$ denote the frame-level CTC posterior for $k\in\Sigma\cup\{\varnothing\}$, where $\Sigma$ is the CTC token vocabulary and $\varnothing$ is the blank symbol. Peak Suppression Decoding (PSD) removes frames satisfying $p_t(\varnothing)>\tau_b$, yielding the ordered retained sequence $\widetilde{\mathbf{p}}=(\widetilde{p}_1,\ldots,\widetilde{p}_R)$, where $R\le T$. Given a candidate $w$, let $\mathbf{y}=(y_1,\ldots,y_L)$ denote its CTC-token sequence. VISH searches for its best constrained CTC alignment over any local interval:
\begin{equation}
m_w=
\max_{\substack{1\le a\le b\le R\\
\boldsymbol{\pi}_{a:b}:B(\boldsymbol{\pi}_{a:b})=\mathbf{y}}}
\sum_{r=a}^{b}\log \widetilde{p}_r(\pi_r),
\qquad
s_w=\frac{m_w}{N_w},
\label{eq:vish_score}
\end{equation}
where $N_w$ is the number of non-blank frames in the maximizing path. Normalizing the accumulated path score by its non-blank support reduces the duration bias of unnormalized path sums. No explicit duration bound is imposed; the constrained CTC states and accumulated posterior score determine the optimal local interval. For the blank-interleaved target $\mathbf{y}'=(\varnothing,y_1,\varnothing,\ldots,y_L,\varnothing)$, the active states follow
\begin{equation}
\alpha_r(s)=\log \widetilde{p}_r(y'_s)
+\max_{q\in\mathcal{P}(s)}\alpha_{r-1}(q),
\label{eq:ctc_viterbi_dp}
\end{equation}
where $\mathcal{P}(s)$ denotes valid CTC predecessor states for stay, step, and skip transitions. Root states are reintroduced after each retained frame, allowing local matches to start throughout the utterance.

Independent candidate-wise scoring repeats these state updates for candidates sharing common token prefixes. M-VISH avoids this redundancy by organizing the candidate list $W=\{w^{(1)},\ldots,w^{(N)}\}$ into a shared-prefix trie, where each edge corresponds to a CTC token. Each active node $u$ maintains a prefix score vector $\alpha_r^u$ and a companion accumulator $\gamma_r^u$ that follows the maximizing path and counts its non-blank aligned frames. The prefix-level confidence is
\begin{equation}
\rho_r(u)=\max_s\alpha_r^u(s).
\label{eq:mv-h-s_priority}
\end{equation}
At each retained frame, only the top-$K_{\mathrm{p}}$ prefixes ranked by $\rho_r(u)$ are retained and expanded. When a terminal node is reached, its best confidence and aligned length update $m_w$ and $N_w$ for the corresponding candidate. Algorithm~\ref{algo:mv-h-s} summarizes the search.

\begin{algorithm}[t]
    \caption{Multi-Word Viterbi-guided Selection of Hotwords~(M-VISH)}
    \label{algo:mv-h-s}
    \normalsize
    \DontPrintSemicolon
    \KwIn{Posteriors $\{p_t\}_{t=1}^{T}$; candidates $W$; prefix budget $K_{\mathrm{p}}$; PSD threshold $\tau_b$.}
    \KwOut{Candidate scores $\{s_w:w\in W\}$.}
    Form $\widetilde{\mathbf{p}}\leftarrow[p_t:p_t(\varnothing)\le\tau_b]$ in temporal order; let $R\leftarrow|\widetilde{\mathbf{p}}|$.\;
    Build a trie for $W$ with a dummy root; set $m_w\leftarrow-\infty$ and $N_w\leftarrow0$.\;
    Initialize active prefixes $P$ from the root children.\;
    \For{$r\leftarrow1$ \KwTo $R$}{
        \ForEach{$u\in P$}{
            Update $\alpha_r^u$ by~\eqref{eq:ctc_viterbi_dp} and update $\gamma_r^u$ along the maximizing predecessors.\;
            $s_u^*\leftarrow\arg\max_s\alpha_r^u(s)$; $\rho_r(u)\leftarrow\alpha_r^u(s_u^*)$.\;
        }
        Retain the top-$K_{\mathrm{p}}$ prefixes in $P$ ranked by $\rho_r(u)$.\;
        \ForEach{terminal $u\in P$ representing $w$}{
            \If{$\rho_r(u)>m_w$}{
                $(m_w,N_w)\leftarrow(\rho_r(u),\gamma_r^u(s_u^*))$\;
            }
        }
        Expand surviving prefixes and initialize child states from their parents.\;
        Add root children to $P$, initialized to start at retained frame $r+1$.\;
    }
    \Return $\{s_w:w\in W\}$, where $s_w=m_w/N_w$ if $N_w>0$, and $s_w=-\infty$ otherwise.\;
\end{algorithm}

A dummy root stabilizes local-match initialization, while PSD suppresses unnecessary propagation through blank-dominated frames. After PSD, independent scoring costs $O(RN\bar{L})$, whereas M-VISH costs $O(RK_{\mathrm{p}}\bar{L})$, with $K_{\mathrm{p}}\ll N$; in the worst case, $R=T$.

\subsection{Bidirectional M-VISH}

\label{sec:fb_v-h-s}

Bi-MVISH supplements forward search when candidate-onset evidence is weak. It reverses both the retained posterior sequence and each candidate CTC-token sequence, constructs the reversed trie, and applies the same M-VISH search. Because CTC collapse commutes with sequence reversal, reversing both sequences preserves valid alignments and transition rules. Mapping the terminal score back to the original candidate gives $s_{w,\mathrm{bwd}}$, which is averaged with the forward score $s_{w,\mathrm{fwd}}$:
\begin{equation}
s_{w,\mathrm{bi}}
=
\frac{1}{2}
\left(
s_{w,\mathrm{fwd}} + s_{w,\mathrm{bwd}}
\right).
\label{eq:bi_mvish}
\end{equation}
M-VISH remains frame-synchronous, whereas Bi-MVISH uses future frames within each segment and therefore trades strict frame-level streaming for stronger segment-level verification.

\section{Experimental Setups}

\subsection{Datasets and bias lists}
\label{sec:dataset}

\textbf{Test datasets.} We use the following test sets:
\begin{itemize}[leftmargin=*,itemsep=1pt,topsep=2pt]
    
    \item \textbf{STOP1/STOP2}~\cite{tomasello2023stop,kong25_interspeech}: English named-entity retrieval sets derived from the STOP test and evaluation splits, containing queries with person names (STOP1) or location names (STOP2). Both are used only for hotword pre-selection evaluation.
    \item \textbf{AISHELL-1-NE}~\cite{shi2024seaco}: A Mandarin named-entity contextual-biasing set derived from AISHELL-1. It is used for both hotword pre-selection and downstream contextual ASR evaluation.
    \item \textbf{Libri-clean/other-2k}~\cite{LibriSpeech}: LibriSpeech test-clean and
    test-other with fixed 2,000-word candidate lists per
    utterance, used for both pre-selection and downstream ASR evaluation; test-other is the more challenging condition.
\end{itemize}

\textbf{Bias lists.} STOP1, STOP2, and AISHELL-1-NE follow their
benchmark protocols. For LibriSpeech, following~\cite{le2021contextualized},
rare words exclude the 5,000 most frequent training words. Each utterance
uses a fixed 2,000-word list containing all ground-truth rare words plus
randomly sampled distractors; the same seeded lists are shared across
methods. For latency analysis, we additionally evaluate list sizes of
100, 500, 1,000, and 2,000.

\subsection{Systems and inference setup}
\label{sec:inference_protocol}
\vspace{-5pt}

We evaluate Whisper-medium~\cite{radford2023whisper} (encoder--decoder ASR) and Prompt-SLAM-ASR~\cite{yang2024ctc} (Speech-LLM contextual ASR). The latter's released model does not support Mandarin; we evaluate it only on LibriSpeech. Frozen SenseVoice-Small~\cite{an2024funaudiollm} supplies VISH's frame-level CTC posteriors and tokenizer for both backbones. We fix $K_{\mathrm p}=40$ and $\tau_b=0.997$ across all datasets and evaluation settings. GLCLAP, M-VISH, and Bi-MVISH share fixed Top-10 selection and the prompt: ``Transcribe speech to text. Some hotwords might help. The hotwords are \{\}''. BR-ASR results are published; the backbone-only setting omits hotwords. For all methods, Table~\ref{tab:retrieval_efficiency} measures candidate-search latency from available speech representations/CTC posteriors to candidate scores/retrieval results, excluding acoustic representation extraction.

\subsection{Evaluation metrics}
\vspace{-5pt}

Best-F1 is the maximum F1 over candidate-score thresholds, measuring target--distractor score separability independently of a deployment selection budget. The fixed Top-10 downstream comparisons instead assess ASR utility under a shared prompt budget. We report character error rate (CER) and biased character error rate (B-CER) on AISHELL-1-NE, and word error rate (WER) and biased word error rate (B-WER) on LibriSpeech~\cite{le2021contextualized}. Biased metrics are computed over reference characters or words in the biasing vocabulary. Lower error rates are better.

\vspace{-5pt}
\section{Results and Analysis}

\subsection{Hotword pre-selection performance}

\label{sec:preselection_results}

\begin{table}[h]
\vspace{-15pt}
\caption{\textbf{Hotword pre-selection performance.}
Best-F1 (\%) across five benchmarks.
A1-NE denotes AISHELL-1-NE; L-clean-2k and L-other-2k
denote LibriSpeech test-clean and test-other with 2,000
candidate hotwords, respectively.
$^*$ denotes our reproduced results; ``--'' indicates
unreported results. Best results are bold.}
\label{tab:hotword_preselection}
\centering
\setlength{\tabcolsep}{1pt}
\begin{tabular}{@{}lccccc@{}}
\toprule
Method & STOP1 & STOP2 & A1-NE & L-clean-2k & L-other-2k \\
\midrule
GLCLAP~\cite{kong25_interspeech}
& -- & -- & 97.0 & -- & -- \\
CLAR~\cite{huang2026clar}
& -- & -- & 97.0 & -- & -- \\
GLCLAP~\cite{kong25_interspeech}$^*$
& 80.5 & 81.2 & 96.6 & 90.5 & 82.4 \\
M-VISH
& \textbf{83.1} & 92.9 & 97.5 & 94.4 & 87.7 \\
Bi-MVISH
& \textbf{83.1} & \textbf{93.8} & \textbf{97.6}
& \textbf{95.3} & \textbf{88.8} \\
\bottomrule
\end{tabular}
\end{table}

CTC-Assistant serves as a representative hypothesis-based CTC
filtering baseline. We use GLCLAP as our primary learned-retrieval
baseline and reproduce it across all five benchmarks. We also
include the original results of GLCLAP and CLAR where available.
CTC-Assistant is omitted from Table~\ref{tab:hotword_preselection}
because its published per-word matching procedure~\cite{yang2024ctc}
does not directly provide compatible full-inventory scores for
threshold-based F1 evaluation. Compared with our reproduced GLCLAP
baseline, M-VISH achieves absolute Best-F1 improvements of
0.9\%--11.7\% across all five benchmarks, with the largest gains
on STOP2 (+11.7\%) and Libri-other-2k (+5.3\%). Bi-MVISH
further improves Best-F1 by up to 1.1\% in absolute terms on
four of the five benchmarks, without additional retrieval-model
training.

\vspace{-5pt}
\subsection{Downstream contextual ASR performance}
\label{sec:downstream_asr}

\begin{table}[t]

\caption{\textbf{Downstream contextual ASR performance.}
WER/B-WER (\%) are reported for LibriSpeech and CER/B-CER (\%)
for AISHELL-1-NE.
GT Bias denotes oracle hotword prompting.
$^*$ denotes our reproduced results.
N/A indicates Prompt-SLAM-ASR model does not support Mandarin.
Best non-oracle results are bold.}
\label{tab:downstream_asr}
\centering
\normalsize
\setlength{\tabcolsep}{2pt}
\renewcommand{\arraystretch}{1}
\begin{tabular}{@{}lccc@{}}
\toprule
ASR System & L-clean-2k & L-other-2k & A1-NE \\
\midrule
Whisper-medium~\cite{radford2023whisper} & & & \\
\quad+No Bias & 3.1/14.0 & 7.1/29.9 & 7.3/21.3 \\
\quad+GT Bias & 2.2/6.2 & 5.6/15.9 & 4.3/7.4 \\
\quad+CTC-Assistant~\cite{yang2024ctc}$^*$ & 2.4/7.8 & 6.2/18.8 & 4.7/9.9 \\
\quad+GLCLAP~\cite{kong25_interspeech}$^*$ & 2.4/7.6 & 6.1/18.2 & 4.6/9.1 \\
\quad+M-VISH & 2.4/7.5 & 6.0/18.3 & \textbf{4.4}/7.8 \\
\quad+Bi-MVISH & \textbf{2.3}/\textbf{6.9} & \textbf{5.9}/\textbf{17.3} & \textbf{4.4}/\textbf{7.6} \\
\midrule
Prompt-SLAM-ASR~\cite{yang2024ctc} & & & \\
\quad+No Bias & 2.0/9.3 & 4.0/20.0 & N/A \\
\quad+GT Bias & 1.2/2.8 & 2.4/5.8 & N/A \\
\quad+BR-ASR~\cite{gong2025br} & 1.4/4.2 & 3.1/9.3 & N/A \\
\quad+CTC-Assistant~\cite{yang2024ctc}$^*$ & \textbf{1.4}/4.4 & 3.2/10.0 & N/A \\
\quad+GLCLAP~\cite{kong25_interspeech}$^*$ & 1.5/4.8 & 3.2/10.7 & N/A \\
\quad+M-VISH & \textbf{1.4}/4.2 & \textbf{3.1}/8.7 & N/A \\
\quad+Bi-MVISH & \textbf{1.4}/\textbf{3.7} & \textbf{3.1}/\textbf{8.0} & N/A \\
\bottomrule
\end{tabular}
\vspace{-15pt}
\end{table}

Table~\ref{tab:downstream_asr} shows improved pre-selection
translates into lower biased error under the same Top-10 prompting
protocol. Compared with reproduced GLCLAP, Bi-MVISH reduces B-WER
from 7.6\% to 6.9\% and from 18.2\% to 17.3\% on Whisper
L-clean-2k and L-other-2k, respectively, while reducing A1-NE B-CER
from 9.1\% to 7.6\%. With Prompt-SLAM-ASR, the corresponding
B-WER decreases from 4.8\% to 3.7\% and from 10.7\% to 8.0\%.
M-VISH is already competitive, while bidirectional verification yields
the most consistent gains without degrading overall WER/CER. The
consistent improvements across both backbones further suggest the
frozen external CTC evidence transfers effectively to downstream
contextual prompting.

\begin{table}[h]
\vspace{-10pt}
\caption{\textbf{Effect of CTC posterior source.}
Comparison of internal and external CTC encoders for Bi-MVISH
with Prompt-SLAM-ASR using 2,000 candidate hotwords per utterance.
Extra denotes additional encoder parameters beyond the ASR backbone.
Clean and Other refer to LibriSpeech test-clean and test-other,
respectively. Best results are bold.}
\label{tab:ctc_source}
\centering
\small
\setlength{\tabcolsep}{2.0pt}
\renewcommand{\arraystretch}{1.05}
\begin{tabular}{@{}cccccc@{}}
\toprule
Encoder & Params. & Extra & Test set & Best-F1 & WER/B-WER \\
\midrule
\multirow{2}{*}{\shortstack{SenseVoice-Small\\(external)}}
& \multirow{2}{*}{234M}
& \multirow{2}{*}{+234M}
& Clean & \textbf{95.3} & \textbf{1.4}/\textbf{3.7} \\
& & &
Other & \textbf{88.8} & 3.1/\textbf{8.0} \\
\midrule
\multirow{2}{*}{\shortstack{WavLM-Large\\(internal)}}
& \multirow{2}{*}{315M}
& \multirow{2}{*}{0}
& Clean & 94.7 & \textbf{1.4}/3.9 \\
& & &
Other & 87.8 & \textbf{3.0}/8.6 \\
\bottomrule
\end{tabular}
\vspace{-15pt}
\end{table}

\subsection{Plug-and-play Adaptability}
\label{sec:ctc_source}

To further validate the plug-and-play property of VISH, we
evaluate Bi-MVISH using either the native CTC posteriors of
Prompt-SLAM-ASR or a frozen external SenseVoice-Small encoder,
without additional training or downstream-model modification.
As shown in Table~\ref{tab:ctc_source}, both sources enable
effective hotword verification. Native posteriors require no
additional encoder or acoustic forward pass, whereas the external
scorer achieves higher Best-F1 and lower B-WER. These results
demonstrate the flexibility of VISH across CTC posterior sources,
with a trade-off between verification accuracy and computational cost.

\subsection{Analysis of bidirectional verification}
\label{sec:bi_mvish_verification}

\begin{figure}[t]
    \centering
    \includegraphics[width=1\linewidth]{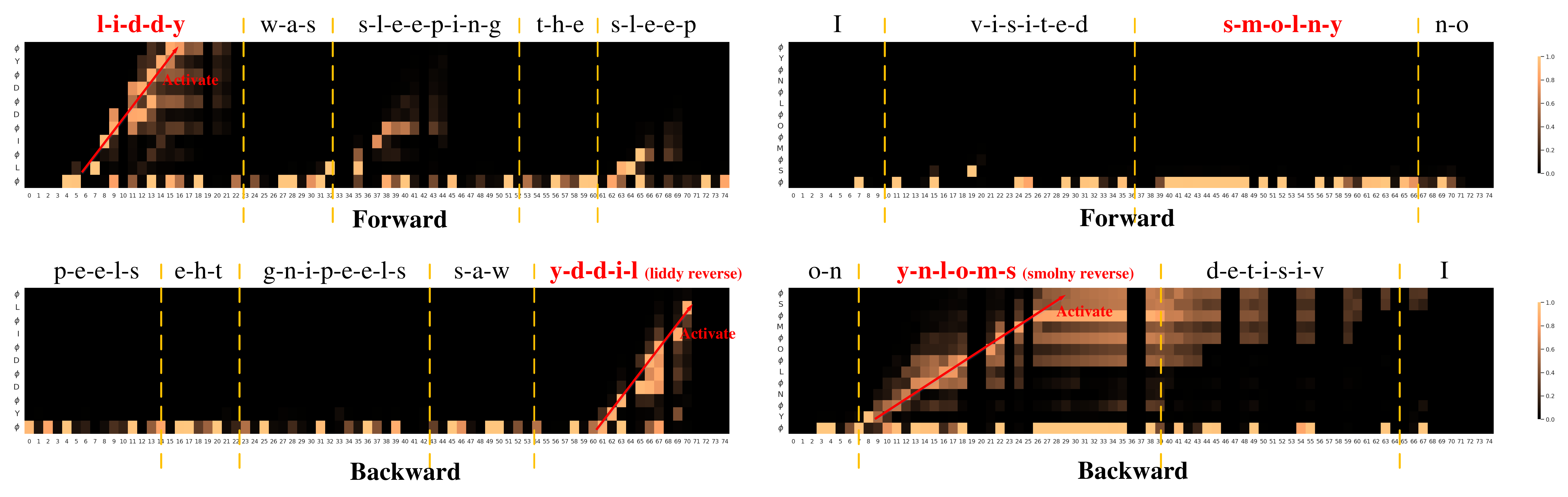}
    \caption{Forward and backward Bi-MVISH alignments for ``liddy'' (left) and ``smolny'' (right). Both directions detect ``liddy'', whereas backward search recovers stronger evidence for ``smolny'' with its weak onset. Yellow dotted lines denote word boundaries.}
    \label{fig:forward_backward}
    \vspace{-10pt}
\end{figure}

Forward Viterbi verification can underestimate a hotword when its onset is acoustically weak or ambiguous. As shown in Figure~\ref{fig:forward_backward}, both directions detect ``liddy'', whereas forward search provides weak evidence for ``smolny'' and backward search anchors on its clearer ending. Averaging the complementary directional scores can therefore compensate for weak forward evidence, providing a qualitative explanation for the additional Bi-MVISH gains in Tables~\ref{tab:hotword_preselection} and~\ref{tab:downstream_asr}.

\begin{table}[h]
\vspace{-10pt}
\caption{\textbf{Candidate-search efficiency.}
Average search latency (ms per utterance) with different bias list
sizes, measured as the number of candidate hotwords per utterance.
Acoustic representation extraction is excluded.}
\label{tab:retrieval_efficiency}
\centering
\setlength{\tabcolsep}{5pt}
\begin{tabular}{@{}lcccc@{}}
\toprule
Bias List Size & 100 & 500 & 1,000 & 2,000 \\
\midrule
CTC-Assistant~\cite{yang2024ctc} & \textbf{46} & 206 & 398 & 817 \\
GLCLAP~\cite{kong25_interspeech} & 61 & \textbf{65} & \textbf{74} & \textbf{78} \\
M-VISH & 71 & 85 & 117 & 203 \\
Bi-MVISH & 131 & 176 & 260 & 400 \\
\bottomrule
\end{tabular}
\vspace{-10pt}
\end{table}

\subsection{Candidate-search efficiency}
\label{sec:retrieval_efficiency}

Table~\ref{tab:retrieval_efficiency} compares the candidate-search
latency of VISH with hypothesis-based filtering and learned
speech--text retrieval methods. All methods are evaluated with
batch size 1 on the same GPU, excluding acoustic representation
extraction. With 2,000 candidate hotwords, M-VISH reduces
search latency from 817~ms for CTC-Assistant to 203~ms,
a relative reduction of 75.2\%. Bi-MVISH achieves a 51.0\%
relative reduction despite performing bidirectional search.
Although GLCLAP achieves the lowest search latency (78~ms),
it yields lower Best-F1 than VISH (Table~\ref{tab:hotword_preselection}).
These results demonstrate that VISH achieves a favorable balance
between hotword pre-selection accuracy and search efficiency
for large candidate inventories.

\section{Conclusion}
\label{sec:conclusion}

We presented VISH, a training-free hotword pre-selector for
candidate-specific verification over CTC posteriors, together with
M-VISH for scalable shared-prefix search and Bi-MVISH for
bidirectional verification. Across five benchmarks, M-VISH achieves absolute Best-F1 improvements of 0.9\%--11.7\% over our reproduced
GLCLAP baseline, while Bi-MVISH
consistently reduces downstream biased error without degrading overall
ASR accuracy. At 2,000 candidates, M-VISH reduces candidate-search
latency from 817 to 203~ms relative to CTC-Assistant. Native CTC
posteriors provide a lower-overhead option when available, while an
external frozen scorer offers stronger plug-and-play verification.

\clearpage
\section{Compliance with Ethical Standards}
\label{sec:ethics}
This study uses existing speech datasets, including STOP,
AISHELL-1-NE, and LibriSpeech. No new human participants were recruited or recorded, and no animal experiments were conducted. All datasets were used in accordance
with their respective access and licensing terms.
\bibliographystyle{IEEEbib-icassp2027}
\bibliography{references}

\end{document}